%% file: main.tex
\documentclass[submission,copyright,creativecommons]{eptcs}
\providecommand{\event}{SOS 2007} 

\usepackage{amsfonts,amsmath,amssymb}
\usepackage{stmaryrd}
\usepackage{mathrsfs}
\usepackage{color}
\usepackage[dvipsnames]{xcolor}
\usepackage{proof}
\usepackage[normalem]{ulem}
\usepackage{newunicodechar}
\usepackage{graphics} 

\usepackage[draft=true, newfloat]{minted}
\usepackage{caption}
\BeforeBeginEnvironment{minted}{\bigskip}
\AfterEndEnvironment{minted}{\bigskip}
\usepackage{etoolbox}
\makeatletter
\AtBeginEnvironment{minted}{\dontdofcolorbox}
\def\dontdofcolorbox{\renewcommand\fcolorbox[4][]{##4}}
\makeatother
\newlength{\fancyvrbtopsep}
\newlength{\fancyvrbpartopsep}
\makeatletter
\FV@AddToHook{\FV@ListParameterHook}{\topsep=\fancyvrbtopsep\partopsep=\fancyvrbpartopsep}
\makeatother
\newminted{coq}{autogobble=true, linenos=false}
\newminted{haskell}{autogobble=true, linenos=false}
\newmintinline{coq}{breaklines, breakafter=_}
\newmintinline{haskell}{}

\newenvironment{code}{\captionsetup{type=listing}}{}
\SetupFloatingEnvironment{listing}{name=Code}

\input{macros}

\usepackage{iftex}

\ifpdf
  \usepackage{underscore}         
  \usepackage[T1]{fontenc}        
\else
  \usepackage{breakurl}           
\fi

\title{Nominal Sets in Rocq}
\author{Fabrício Sanches Paranhos
\institute{Nethermind\\ London, UK}
\email{fabricio@nethermind.io}
\and
Daniel Ventura
\institute{Univ. Federal de Goiás\\
Goiânia, Brazil}
\email{ventura@ufg.br}
}
\def\titlerunning{Nominal Sets in Rocq}
\def\authorrunning{F.S. Paranhos \& D. Ventura}
\begin{document}
\maketitle

\begin{abstract}
Nominal techniques have been praised for their ability to formalize grammars with binding structures closer to their informal developments. At its core, there lies the definition of nominal sets, which capture the notion of name (in)dependence through a simple, and uniform, metatheory based on name permutations. We present a formal constructive development of nominal sets in Rocq (formerly known as Coq), with its main design and project decisions. Furthermore, we formalize the concepts of freshness, nominal $\alpha$-equivalence, name abstraction, and finitely supported functions. Our implementation relies on a type class hierarchy which, combined with Rocq generalized rewriting mechanism, achieves concise definitions and proofs, whilst easing the well-known ``setoid hell'' scenario. We conclude with a discussion on how to obtain the constructive $\alpha$-structural recursion and induction combinators, towards a nominal framework.
\end{abstract}

\input{intro}
\input{const-nominal}
\input{rocq-development}
\input{discussion-related}

%
%
%
\bibliographystyle{eptcs}
\bibliography{ref}

\end{document}

%% file: macros.tex
\long\def\ignore#1{\relax}

\newcommand{\deft}[1]{\textbf{#1}}

\newtheorem{theorem}{Theorem}[section]

\newunicodechar{λ}{$\lambda$}
\newunicodechar{≠}{$\neq$}
\newunicodechar{≈}{$\approx$}
\newunicodechar{α}{$\alpha$}
\newunicodechar{∙}{$\bullet$}
\newunicodechar{∀}{$\forall$}
\newunicodechar{∃}{$\exists$}
\newunicodechar{∧}{$\land$}
\newunicodechar{∨}{$\lor$}
\newunicodechar{≡}{$\equiv$}
\newunicodechar{ɛ}{$\varepsilon$}
\newunicodechar{∉}{$\notin$}
\newunicodechar{⊆}{$\subseteq$}
\newunicodechar{ₑ}{$_{e}$}
\newunicodechar{ₐ}{$_{a}$}
\newunicodechar{ₛ}{$_{s}$}
\newunicodechar{↔}{$\leftrightarrow$}
\newunicodechar{⟹}{$\Rightarrow$}
\newunicodechar{∖}{$\setminus$}
\newunicodechar{Α}{$\mathbb{A}$}
\newunicodechar{⟦}{$\llbracket$}
\newunicodechar{⟧}{$\rrbracket$}
\newunicodechar{∪}{$\cup$}
\newcommand\new[0]{\reflectbox{\ensuremath{\mathsf{N}}}}

\newcommand{\eg}{\textsl{e.g.}}
\newcommand{\ie}{\textsl{i.e.}}

\newcommand{\lvec}{\langle}
\newcommand{\rvec}{\rangle}

\newcommand{\nel}{\varepsilon}

\newcommand{\aeq}{\approx_{\alpha}}

\newcommand{\nameset}{\mathbb{A}}
\newcommand{\permset}{P_{\mathbb{A}}}
\newcommand{\naset}[1]{[\nameset]#1}

\newcommand{\action}{\bullet}
\newcommand{\neutral}{\varepsilon}
\newcommand{\BB}{\mathbb{B}}

\newcommand{\supp}{\mathtt{supp}}
\newcommand{\nabs}[2]{[#1] #2}
\newcommand{\tofs}{\to_{\mathtt{fs}}}

\newcommand{\id}{\mathtt{id}}
\newcommand{\nil}{\mathtt{nil}}

\newcommand{\fresh}[1]{\mathtt{fresh}(#1)}

%% file: intro.tex
\section{Introduction}

Nominal techniques \cite{Pitts2016} have been praised due to its capacity to formalize syntax with binders closer to their informal developments (pen and paper proofs) \cite{Bengtson2007,CopelloST21}. However, given the elaborate metatheory of nominal sets \cite{Pitts2009} and the lack of a formal development of an infrastructure (cf. \cite{Urban2008}), many nominal-based formalization \cite{Ambal2020,AyalaSFN19,AyalaSFS21,Aydemir2007,Copello2018,CopelloST21,Copello2016,Qinxiang2023}, named pragmatic here, choose to apply the bare minimum ideas from nominal logic to a specific issue, \eg\ in a formalization of the $\lambda$-calculus and the $\pi$-calculus.

The consequence of a pragmatic approach is that, whenever a new development chooses to use nominal logic, the whole infrastructure needs to be rebuilt. Nominal frameworks, like Isabelle's nominal package \cite{Urban2008}, provide the necessary infrastructure and automation, which can reduce the amount of man-hour considerably. Unfortunately, to the best of our knowledge, no other proof assistant has such a framework. In particular, there is a proof of concept in Rocq (formerly known as Coq) \cite{Aydemir2007}, with no further development, that presents an approach based on axiomatization. In \cite{Aceto19}, Pitts mentions the lack of support for nominal techniques in dependent-type based systems, such as Rocq and Agda, whilst posing a question about the possibility of a dependent type theory corresponding, under the Curry-Howard Correspondence, to a constructive nominal logic that could be useful as a nominal framework. In \cite{Choudhury2015} such a possibility is investigated in an Agda formal development, which relies heavily on setoids, and formalizes  the core aspects of the nominal theory: support, freshness and name abstraction.

We propose a formalization of constructive nominal sets in Rocq, a foundation towards a nominal framework, following the approach in \cite{Choudhury2015} closely. Nominal sets modelled as a typeclass hierarchy was already introduced in \cite{PV2022} and, besides formalization of the core of the nominal theory, we specified two notions related to the derivation of $\alpha$-structural principles: the freshness theorem and the freshness condition for binders (FCB). We claim that by using Rocq's typeclass \cite{Sozeau2008} and the generalized rewriting mechanism \cite{Sozeau2009}, we achieve much shorter and simpler proofs, with some of them close resembling its informal, pen and paper, counterpart.

The remaining text is structured as follows: in Section~\ref{sec:constructive-nominal} we introduce the (constructive) nominal sets and related notions; in Section~\ref{sec:rocq-devel} we discuss some aspects of the Rocq specification; and in Section~\ref{sec:discussion-related} we conclude with related and future work.

Full development, using Rocq version 8.20.1, can be found at \url{https://github.com/fasapa/nominal/tree/lsfa2025}.

%% file: const-nominal.tex
\section{Constructive Nominal Sets}\label{sec:constructive-nominal}

Nominal sets capture the notion of name (in)dependence through a simple, and uniform, metatheory based on name permutations. The \deft{set of names} $\nameset$ is an infinite set of \deft{atoms}, where the internal structure of its terms is irrelevant and the only relevant attribute is the identity relation. By enriching a structure built from $\nameset$ with (left group) action of permutations such a notion of name dependence can be defined, where the set of names that a structure depends on is called \deft{support} while its complement is formed by the so-called \deft{fresh names}. In other words, nominal sets give us the tools to express name dependence for any of its elements by means of name permutation only. In what follows, we present the key notions in order to define nominal sets (see \cite{Pitts2009} for further details).

Given any set $A$, the set $\mathcal{S}_A$ of bijections $f : A \to A$ is a well-known algebraic structure, known as the \deft{permutation group}, when combined with function composition $\circ$. In particular, when $A$ is infinite, the subset $P_A \subset \mathcal{S}_A$ formed only by $f$ such that $\{x \in A \,|\, x \neq f(x) \}$ is finite, called \deft{finite permutations}, is also a group with the same operation. $\lvec \permset, \circ \rvec$ is called the \deft{name permutation group}, where $p^{-1}$ denotes the inverse of $p \in \permset$ and $\id$ its neutral element. Another well-known result about finite permutations\footnote{Note that for $A$ a finite set, any permutation in $\mathcal{S}_A$ is finite.} is that any permutation may be represented as a composition of \deft{transpositions} of the form $(a ~ b)$, also called \deft{swapping}, mapping $a$ to $b$ and $b$ to $a$ while any other $c \in A$ is mapped to itself. Finite permutations represented as sequences of swaps is the base of the current formal development (see Sec.~\ref{sec:rocq-devel}).

Let $X$ be a set and $\lvec G, * \rvec$ be a group with neutral element $\nel$. \deft{Left action} of $G$ on $X$ is a function $(\action) : G \times X \rightarrow X$ satisfying:
\[ (\forall x \in X),\; \neutral \action x = x  \qquad\qquad (\forall g,\,h \in G) (\forall x \in X),\; g \action (h \action x) = (h * g) \action x \]
Such an action is called a \deft{permutation action} when $G = \permset$ and $X$ is then called a \deft{Perm-set}.

Let $X$ be any Perm-set, we say that a finite $S \subset \nameset$ is a (finite) \deft{support} of some $x \in X$ if:  
\begin{equation}\label{def:fsupp}	
(\forall p \in \permset), (((\forall a \in S), p \action a = a) \Rightarrow p \action x = x)
\end{equation}
A support is then a finite collection of all names in which an element $x$ depends on, in the sense that any permutation fixing each name in this set also fixes $x$. A Perm-set $X$ is a \deft{nominal set} if all of its elements have a finite support. For instance, $\nameset$ is itself a nominal set --where $a$ is supported by $\{a\}$, for any $a \in \nameset$-- and the boolean set $\BB$ is the so-called \deft{trivial nominal set} --where each element is supported by the empty set. From now on, $X$ denotes a nominal set unless otherwise stated. 

Supports are not unique and, given $S_1$ and $S_2$, two supports of some $x \in X$, $S_1 \cap S_2$ is also a support of the same element \cite{Pitts2009}. We can thus define the \deft{least support} of an $x \in X$ as follows:
\begin{equation}\label{def:supp}
\supp(x) \triangleq \bigcap \left\{ S \mid S \text{ supports } x \right\}
\end{equation}
In other words, the least support of an element $x$ is the all and only names in which $x$ depends on. In a classical setting, every finitely supported element is proven to have a least support \cite{Pitts2009}. Least supports allows to obtain a stronger statement than (\ref{def:fsupp}):
\begin{equation}\label{def:lsupp}	
(\forall p \in \permset), (((\forall a \in \supp(x)), p \action a = a) \iff p \action x = x)
\end{equation}
\deft{Freshness} relation is then defined as the complementary notion of least support, where given $X$ and $Y$, nominal sets, for any $x \in X$, $y \in Y$: 
\begin{equation}\label{def:fresh}
x~\#~y \Leftrightarrow \supp(x) \cap \supp(y) = \emptyset
\end{equation}
In particular, if $X = \nameset$ then $x~\#~y \Leftrightarrow x \notin \supp(y)$ and $x$ is said to be \deft{fresh} for $y$. Given $y$ finitely supported, the set $\nameset \smallsetminus \supp(y)$ of fresh names for y is infinite, thus to choose a fresh name is always possible, called \deft{choose-a-fresh-name principle}. Freshness is an \deft{equivariant} relation, \ie\ closed for permutation action:
$\forall p \in P_\nameset, x~\#~y \Rightarrow (p \action x)~\#~(p \action y)$. In particular, equivariance can be restated when $X = \nameset$ as: 
\begin{equation}\label{def:fresh-inv}
\forall p \in P_\nameset,\; x\notin\supp(y) ~\Rightarrow~ (p \action x) \notin \supp(p \action y) 
\end{equation}
It is worth noticing that support (and freshness) is relative to the equality considered in $X$. For instance, let $X$ be the set $\Lambda$ of $\lambda$-terms and $t \in \Lambda$. The support of $t$ is the set of all the term-variables occurring in $t$, when the syntactic equality in $\Lambda$ is considered, while it is the set of free variables when $\alpha$-equivalence, with its associated equality on $\Lambda$, is considered instead.

Two key notions can be defined in a rather abstract way for nominal sets: $\alpha$-equivalence and name abstraction. First, 
$\mathbf{\alpha}$-{\bf equivalence} is a binary relation on $\nameset \times X$, denoted by $\aeq$, defined by:
\begin{equation}\label{def:e-alfa}
(a, x) \aeq (b, y) ~\triangleq~ (\exists c \in \nameset),\; c ~\#~ (a, b, x, y) ~\land~ (a~c) \action x = (b~c) \action y
\end{equation}
where $c ~\#~ (a, b, x, y)$ denotes $c ~\#~ a \land c ~\#~ b \land c ~\#~ x \land c ~\#~ y$. By the Some/Any Theorem (Thm. 3.9 in \cite{Pitts2009}) an equivalent definition can be given:
\begin{equation}\label{def:a-alfa}
(a, x) \aeq (b, y) ~\triangleq~ (\forall c \in \nameset),\; c ~\#~ (a, b, x, y) ~\Rightarrow~ (a~c) \action x = (b~c) \action y
\end{equation}
Such equivalences are considered in \cite{Pitts2009} with respect to the $\new$ quantifier, where $(\new a \in \nameset)\varphi$ means that $\varphi$ holds for all but finitely many atoms. Changing from ``some fresh'' to ``any fresh'' name is identified as a common pattern while reasoning with fresh names \cite{Pitts2009}, and the utility of having two such definitions is briefly discussed in the locally nameless approach \cite{Chargueraud12}, where a cofinite quantification is proposed instead\footnote{Cofinite quantification restricted to finitely supported predicates corresponds to the $\new$ quantification \cite{Pitts23}.}. We call \deft{some/any principle} the reasoning in proving equivalences between existential/universal predicates in the formal development.

Second, the \deft{name abstraction set} of $X$ is the set $\nameset \times X$ modulo $\aeq$, denoted by $\naset X$. The name abstraction $\nabs a x$ denotes the class $c \in \naset X$ where $(a,x) \in c$. Observe that permutation action on $\naset X$ is well-defined by $p \action \nabs a x \triangleq \nabs{p \action a}{(p \action x)}$ and is proved that $\supp(\nabs a x) = \supp(x) \smallsetminus \{a\}$, for any $\nabs a x \in \naset X$. The proof that $\naset X$ is a nominal set relies on that a quotient of a nominal set by an equivariant equivalence relation is also nominal \cite{Pitts2009}$: \nameset \times X$ is nominal --product of two nominal sets-- and $\aeq$ is proved to be equivariant.

Given $X,Y$ nominal sets, let $X \to Y$ be the set of all functions from $X$ to $Y$. Considering the permutation action defined by $p \action F \triangleq \lambda x \in X \to p \action (F(p^{-1} \action x))$, the set: 
\[X \tofs Y ~\triangleq~ \{ F \in X \to Y \,\mid\, (\exists S \subset \nameset)\mbox{, $F$ is supported by finite $S$} \}\] 
of the \deft{finitely supported functions} is equivariant\footnote{If $F$ is supported by $S$ then $p \action F$ is supported by $\{p(a) \mid a \in S\}$.} and is a nominal set, called \deft{nominal function set.} Finitely supported functions are essential to obtain $\alpha$-structural recursion and induction principles, as presented in \cite{Pitts2006}. Two last properties are needed to obtain a recursion combinator: the Freshness theorem and the Freshness Condition for Binders (FCB).

\begin{theorem}[Freshness theorem \cite{Pitts2006}]\label{thm:fresh-theorem} 
Given a nominal set $X$ and a finitely supported function $h \in \nameset \tofs X$
satisfying
\[(\exists a \in \nameset), a ~\#~ h \land a ~\#~ h(a)\] then exists a unique element $\fresh h \in X$ such that
\[(\forall a \in \nameset), a ~\#~h \Rightarrow h(a) = \fresh h\]
\end{theorem}

The theorem in \cite{Pitts2009} is stated considering  partial functions and $\new$ quantification over the set of names (see Theorem 3.11, \cite{Pitts2009}). A function $f: X \rightarrow Y$ is \deft{respectful} to $\approx_X$ and $\approx_Y$, equalities for $X$ and $Y$ respectively, if \[(\forall a~b \in X),~ a \approx_X b \Rightarrow f(a) \approx_Y f(b)\] We say that a function \deft{respect} equivalence relations if it is respectful to their induced equalities.

\begin{theorem}[Freshness Condition for Binders (FCB) \cite{Pitts2009,Choudhury2015}]\label{thm:fcb} 
Given nominal sets $X$ and $Y$ and a finitely supported function $f \in (\nameset \times X) \tofs Y$
satisfying
\[(\exists a \in \nameset), a ~\#~ f \land (\forall x \in X), a ~\#~ f(a, x)\] then exists a unique function $\overline{f} \in [\nameset] X \tofs Y$ such that
\[(\forall a \in \nameset), a ~\#~ f \Rightarrow (\forall x \in X), \overline{f}([a]x) = f(a, x)\]
\end{theorem}

Any function related to some binder would have the signature $\nameset \times X \tofs Y$ and it would need to respect the nominal $\alpha$-equivalence. Once such a compatibility is proved, we can lift the argument $\nameset \times X$ into $[\nameset] X$, ensuring the function is well-defined over the $\alpha$-equivalence classes. FCB provides a simpler alternative, based on freshness, to this otherwise difficult  approach. With FCB we can construct a well-defined function over the $\alpha$-equivalence classes, with type $[\nameset]X \tofs Y$, from a simpler one, with type $\nameset \times X \tofs Y$, where a some/any principal is helpful since we would just need to show that the condition is valid for a single name.

We postpone a discussion on the $\alpha$-structural recursion and induction, including the related some/any principle of freshness condition for binders (FCB), to Section~\ref{sec:discussion-related}.

\paragraph{Constructive Nominal Sets:} Existence of least supports cannot be proved constructively \cite{Swan2016,Swan2017}. In fact, admitting the existence of least supports is equivalent to the non-constructive \emph{weak limited principle of omniscience} (see \cite{Swan2017} for a discussion). In constructive developments such as \cite{Choudhury2015, Pagano2023} the least support is replaced by \deft{``some'' support}. In other words, given $x \in X$, $S_x$ is any set satisfying (\ref{def:fsupp}) which can also be seen as an upper bound for $\supp (x)$ since $\supp(x) \subseteq S_x$, for any such a set. A characterisation of support based on swaps is given in \cite{Pitts2009}, and used in \cite{Choudhury2015}, where $S_x$ is any $S$ satisfying:
\begin{equation}\label{def:swap-supp}	
(\forall a,b \in \nameset \smallsetminus S), (a~b) \action x = x
\end{equation}

Notions such as freshness need to be addressed while considering some instead of least support. For instance, $a \notin S_x \Rightarrow a ~\#~ x$ for any $S_x$ but $a ~\#~ x \not\Rightarrow a \notin S_x$, since $S_x$ is not unique. Therefore, equivariance of the freshness relation cannot be stated as in (\ref{def:fresh-inv})  since $a \notin S_x ~\not\Rightarrow~ (p \action a) \notin S_{(p \action x)} $, for some $p \in P_\nameset$. \deft{Constructive freshness} is then defined in \cite{Choudhury2015} by:
\begin{equation}\label{def:c-fresh}
 a ~\#~ x ~\triangleq~ (\exists b \in \nameset),\; b \notin S_x \land (a~b) \action x = x
\end{equation}
and a dual definition, proved to be equivalent by a some/any principle:
\begin{equation}\label{def:c-fresh-a}
 a ~\#~ x ~\triangleq~ (\forall b \in \nameset),\; b \notin S_x \Rightarrow (a~b) \action x = x
\end{equation}
The relation is proved to be equivariant \cite{Choudhury2015} and $\alpha$-equivalence can then be defined as in (\ref{def:e-alfa}-\ref{def:a-alfa}). It is worth noticing that a Some/Any Theorem cannot be proved in the current development\footnote{One would need to prove that \coqinline{Prop} is a nominal set, demanding a nominal calculus of inductive constructions.} thus equivalences between universal and existential formulas are proved instead, referred to as \deft{some/any principles}. They are especially important for a nominal framework, since these some/any principles allow us to generalize a proof of a property about a single name to any name. Proofs of such a principle in \cite{Choudhury2015} follows the reasoning in Lemma 3.7 from \cite{Pitts2009}, used to prove the Some/Any Theorem. The present development follows the same approach. Name abstraction is defined as before, with a different proof (see Section~\ref{sec:rocq-devel}).

When considering syntactic equality as the equivalence relation, the formalization of the nominal function set demands the axiom of functional extensionality. To avoid adding the extensionality axiom, Choudhury worked with setoids in \cite{Choudhury2015}, which quickly became, in his words, ``unwieldy''. To escape the ``setoid hell'' scenario \cite{Altenkirch2017}, we leverage Rocq's typeclass together with typeclass based generalized rewriting mechanism, and model nominal sets as an unbundled typeclass hierarchy \cite{PV2022,Spitters2011}.

Last but not least, the Freshness Theorem~\ref{thm:fresh-theorem}, not formalized in \cite{Choudhury2015}, was proved in the current development, with an equivalent handy version also proved/used in the Rocq specification:
\begin{equation}\label{def:freshness-handy}
(\forall a,\,b \in \nameset), ((\exists c \in \nameset), c \notin S_h \land c ~\#~ h(c)) \land a,b ~\#~ h \Rightarrow h(a) = h(b)
\end{equation}

%% file: rocq-development.tex
\section{Rocq Development}\label{sec:rocq-devel}

In previous work \cite{PV2022}, we proposed a typeclass hierarchy, with the bundle/unbundle approach \cite{Spitters2011}, to formalize nominal sets in a constructive setting. It follows closely \cite{Choudhury2015}, a category theory oriented formalization in Agda. A particular difference in our development is the heavy employment of Rocq typeclass based generalized rewriting to mitigate the setoid burden reported in \cite{Choudhury2015}.

In this section, we present key aspects in the current approach. In particular, further developments since \cite{PV2022}, such as the addition of the concepts of freshness, name abstraction, supported functions and related properties, \eg\ the Freshness Theorem, not formalized in \cite{Choudhury2015}. Remark that permutations are considered for a set of names $\nameset$. In some places, we use \coqinline{(* ... *)} to indicate that part of the code was omitted for brevity, highlighting parts of code presented explicitly.

\paragraph{Names} The set of atoms is defined as an abstract data type satisfying two properties: infiniteness and decidable equality of its elements. Defined using the Rocq module system\footnote{Modules in Rocq/OCaML are similar to interfaces in Java.}, the set of natural numbers is used to implement the module (Code \ref{code:names}).

\begin{code}
\begin{minipage}[t]{0.5\linewidth}
\begin{coqcode}
Module Type ATOMIC.
  Parameter t : Set.
  Axiom dec : EqDecision t.
  Axiom inf : Infinite t.
End ATOMIC.
\end{coqcode}
\end{minipage}
\hspace{-5em}
\begin{minipage}[t]{0.5\linewidth}
\begin{coqcode}
Module Atom: ATOMIC.
  Definition t := nat.
  Instance dec : EqDecision t := Nat.eq_dec.
  Instance inf : Infinite t := nat_infinite.
End Atom.
\end{coqcode}
\end{minipage}
\begin{coqcode}
Notation Name := Atom.t.
Definition NameSet := (listset Name).
\end{coqcode}
\caption{Names}
\label{code:names}
\end{code}
The infinite axiom for names is necessary to implement the choose-a-fresh-name principle (see Section~\ref{sec:constructive-nominal}). From a proof of infinite for a type \coqinline{X}, \eg, \coqinline{NameSet} in Code~\ref{code:names}, the coq-std++ library\footnote{\url{https://plv.mpi-sws.org/coqdoc/stdpp/}}, an extended standard Rocq library developed by the same authors of the Iris project \cite{Jung2015}, makes available a lemma \coqinline{exist_fresh: ∃ x, x ∉ X}.

\paragraph{Permutations} We follow \cite{Choudhury2015,Urban2008} and specify name permutations as a list of swaps with operation \mintinline{coq}{perm_swap} that applies the permutation to a name (Code~\ref{code:swap-perm}). Every such a permutation in \coqinline{Perm} is then finite and computation of composition and inverse are trivial, corresponding to list concatenation and reverse, respectively. For the sake of comparison, a representation based on an abstract notion of bijections would need an extra property, addressing finiteness, and the existence of inverses in the general case would need the axiom of choice.

\begin{code}
\begin{coqcode}
Definition Swap: Type := (Name × Name).
Definition swap ('(a,b): Swap) (c: Name): Name :=
  if decide (a = c) then b else if decide (b = c) then a else c.
Notation "⟨ a , b ⟩" := (@cons Swap (a,b) nil).
Notation Perm := (list Swap).
Definition perm_swap (p: Perm) (a: Name): Name := 
  foldl (fun x y => swap y x) a p.
\end{coqcode}
\caption{Swap and Permutation.}
\label{code:swap-perm}
\end{code}

Such a representation forms a group, with concatenation as the binary operator ($+$), empty list as the neutral element ($\varepsilon$) and inverse function as the list reverse function ($-$). Permutation representation is not unique though, \eg\ $[(a,a)]$ and $\nil$ (empty list) both correspond to the identity function $\id$. We consider them as a setoid instead \cite{BartheCP03}, \ie\ as the set of lists of swaps modulo an equivalence relation for permutations, corresponding to an instance of functional extensionality which is provable in the constructive setting. Besides defining an instance of the typeclass \coqinline{Equivalence} --by providing an equivalence relation--, we need to provide instances of the \coqinline{Proper} typeclass --proving that a function preserves the equivalence relations of its domain and codomain--, to enable the rewriting of equivalent terms when defining a setoid in Rocq. For instance, considering the binary operator of some group \coqinline{A} and an equivalence relation over the same group, denoted by \coqinline{≡@{A}}, \coqinline{Proper} is defined as:

\begin{coqcode}
grp_op_proper: Proper ((≡@{A}) ⟹ (≡@{A}) ⟹ (≡@{A})) (+)
\end{coqcode}
which unfolds to a proof stating that if the arguments of the operator are equivalent, then so must be their results. We remark that, in some cases, Rocq could not find the correct relations to rewrite, so we opted to bundle equivalent and proper relations into their respective typeclasses (\eg\ Code \ref{fig:perm-type} below).

\paragraph{Nominal Sets} Perm-set is defined as the typeclass \coqinline{PermT}, with a permutation action over a set \coqinline{X}~(Code \ref{fig:perm-type}). The carrier set \coqinline{X} is a setoid (\coqinline{gact_setoid}) thus the permutation action needs to respect equivalences on both \coqinline{Perm} and \coqinline{X} sets, realised by the \coqinline{Proper} typeclass (\coqinline{gact_proper}).

\begin{code}
\begin{coqcode}
Class PermAction (X: Type) := action: Perm → X → X.
Infix "•" := action.

Class PermT (X: Type) `{PermAction X, Equiv X}: Prop := {
  gact_setoid :> Equivalence(≡@{X});
  gact_proper :> Proper ((≡@{Perm}) ==> (≡@{X}) ==> (≡@{X})) (•);

  gact_id: ∀ (x: X), ɛ • x ≡@{X} x;
  gact_compat: ∀ (p q: Perm) (x: X), p • (q • x) ≡@{X} (q + p) • x
}.
\end{coqcode}
\caption{(Name) Permutation Action}
\label{fig:perm-type}
\end{code}

The \coqinline{Nominal} typeclass is a direct extension of \coqinline{PermT}, since nominal sets are Perm-sets such that all members are finitely supported, as indicated by the \coqinline{:>} operator defining the \coqinline{nperm} field as a coercion from \coqinline{Nominal} to \coqinline{PermT} (Code \ref{code:nominal}). Moreover, the \coqinline{support_spec} field corresponds to the property that any \coqinline{x} in \coqinline{X} is supported by \coqinline{(support x)}, \ie\ the \coqinline{support} function maps each \coqinline{x} to some support $S_x$, as characterised by (\ref{def:swap-supp}).

\begin{code}
\begin{coqcode}
Class Support X := support: X → NameSet.

Class Nominal (X: Type) `{Support X}: Prop := {
    nperm :> PermT X;
    support_spec: ∀ (x: X) (a b: Name),
        a ∉ support x → b ∉ support x → ⟨a,b⟩•x ≡@{X} x
}. 
\end{coqcode}
\caption{Nominal Set}
\label{code:nominal}
\end{code}

A protocol to define a nominal set $N$ consists of five steps: (1) define a permutation action on $N$, (2) define a support function, (3) define and prove an equivalence relation for the elements of $N$, (4) prove an instance of \coqinline{PermT} with $N$ as the carrier set, (5) prove an instance of \coqinline{Nominal} for $N$.

\paragraph{Freshness} Constructive freshness is defined as in (\ref{def:c-fresh}), where an equivalent definition, proved by the some/any principle \mintinline{coq}{fresh_some_any_iff}, is also presented (see Code \ref{code:freshness}).

\begin{code}
\begin{coqcode}
Definition freshP_e `{Nominal X} (a: Name) (x: X): Prop := 
   ∃ (b: Name), b ∉ support x ∧ ⟨a,b⟩ • x ≡@{X} x.
   
Definition freshP_a `{Nominal X} (a: Name) (x: X): Prop := 
   ∀ (b: Name), b ∉ support x → ⟨a,b⟩ • x ≡@{X} x.

Infix "#" := freshP_e. Infix "#ₐ" := freshP_a.

Lemma fresh_some_any_iff `{Nominal X} (a: Name) (x: X): a # x ↔ a #ₐ x.
\end{coqcode}
\caption{Constructive Freshness + Some/Any Freshness Lemma}
\label{code:freshness}
\end{code}

\paragraph{Name Abstraction} Our first implementation of the name abstraction set was a direct adaptation from Agda \cite{Choudhury2015} into Rocq, \ie\ as a pair of name and term, with the corresponding equivalence relation to represent $\approx_\alpha$ (see Section~\ref{sec:constructive-nominal}). The usage of name/term pairs turned out to be problematic while working with supported functions that also had another set of name/term pairs as domain. Our definition of supported functions are uncurried, thus $n$-ary functions have their domain represented by $n$-tuples. Consequently, the typeclass instance search algorithm assigned $\approx_\alpha$ for any function with such a domain, even when there is no equivalence relation defined for the set of pairs. A new Record type called \coqinline{NameAbstraction}, used to represent name abstraction, solved the issue. In other words, instead of defining the equivalence relation over sets of the form $\nameset \times X$, $X$ a nominal set, it is defined over this new type, avoiding the assignment of a $\approx_\alpha$ relation by the typeclass instance search algorithm for such a pair with no equivalence relation defined over the set. As a result, the nominal $\alpha$-equivalence ($\approx_\alpha$) is defined as a relation over the \coqinline{NameAbstraction} type.

\begin{code}
\begin{coqcode}
Record NameAbstraction `{Nominal X} := mkAbstraction { name: Name; term: X }.
Notation " '[Α]' X " := (NameAbstraction X).
Notation " ⟦ a ⟧ x " := ({| name := a; term := x |}).

Instance alpha_equiv_e `{Nominal X}: Equiv [Α]X | 0 :=
   λ x y, ∃ (c: Name), c #[x.(name), y.(name), x.(term), y.(term)] ∧ 
   ⟨c,x.(name)⟩ • x.(term) ≡@{X} ⟨c,y.(name)⟩ • y.(term).

Instance alpha_equiv_a `{Nominal X}: Equiv [Α]X | 1 := 
   λ x y, ∀ (c: Name), c #[x.(name), y.(name), x.(term), y.(term)] → 
   ⟨c,x.(name)⟩ • x.(term) ≡@{X} ⟨c,y.(name)⟩ • y.(term).

Infix "≈α" := (alpha_equiv_e). Infix "≈αₐ" := (alpha_equiv_a).

Lemma alpha_some_any `{Nominal X} (a b: Name) (x y: X): 
   ⟦a⟧x ≈α ⟦b⟧y ↔ ⟦a⟧x ≈αₐ ⟦b⟧y.

Instance `{Equiv X}: Equiv [Α]X := (≈α).
Instance `{PermT X}: PermT [Α]X. Proof. (* ... *) Qed.
Instance `{Nominal X}: Nominal [Α]X. Proof. (* ... *) Qed.

Lemma nabs_action `{Nominal X} p a (x: X): p • (⟦a⟧x) = ⟦p • a⟧(p • x).

Lemma nabs_support `{Nominal X} a (x: X): support ⟦a⟧x = support x ∖ {[a]}.
\end{coqcode}
\caption{Name Abstraction}
\label{code:name-abs}
\end{code}
Proving that name abstraction is a nominal set follows pretty much the general protocol mentioned above. It is worth noticing that, once a some/any principle for name abstraction is proved, there are two possible instances for the name abstraction equivalence relation. 
Unlike Haskell, that only allows one instance per class per type, Rocq typeclasses does not have this restriction. So, Rocq provides a mechanism to prioritize one instance over another during instance search. The notation \coqinline{Typeclass | n} indicates a priority level, with \coqinline{n} a natural number, with $0$ indicating the highest priority. Both permutation action and support for name abstraction are presented as lemmas in Code~\ref{code:name-abs}, making it comparable with the definitions from Section~\ref{sec:constructive-nominal} thus improving the readability.

\paragraph{Finitely supported functions} The definition of the nominal set of supported functions poses particular challenges. Early in the development, we approached them with full unbundle, where the record \coqinline{FunSupp} becomes a \coqinline{Class} which is parametrized by all its members, showing an instance of the \coqinline{Nominal} typeclass for each function separately. A major drawback in the approach was the impossibility to derive an instance for a composition of two nominal functions from their instances. And so, we opted out for a bundled alternative, similar to how bundled morphisms are defined in the Lean MathLib \cite{Baanen2024}. Supported functions are records parametrized by their domain and codomain.
The advantage of this technique is that, we can bundle the support and the respective properties necessary to prove that it is nominal and that it respects the domain and codomain equivalences. On the other side, a lot of extra code is necessary to make it behave as regular Rocq functions, \ie\ that can be applied, computed and composed with other functions defined in the system.

\begin{code}
\begin{coqcode}
Context (X Y: Type) `{Nominal X, Nominal Y}.
Record FunSupp: Type := mkFunSupp {
  f_car :> X → Y;
  f_supp: NameSet; (* Function support *)
  f_proper: Proper ((≡@{X}) ⟹ (≡@{Y})) f_car;
  f_supp_spec: ∀ (a b: Name), a ∉ f_supp → b ∉ f_supp →
      ∀ (x: X), (⟨a,b⟩•(f_car (⟨a,b⟩•x))) ≡@{Y} f_car x
}.

Notation " A '→ₛ' B " := (FunSupp A B).
Notation "'λₛ' x .. y , t" := (* ... *).
Notation "'λₛ[' S ']' x .. y , t" := (* ... *).

Context `{Nominal X, Nominal Y}.
Instance: Equiv (X →ₛ Y) := fun f g => ∀ (x: X), f x ≡@{Y} g x.
Instance: PermAction (X →ₛ Y) := 
                       fun (p: Perm) (f: X →ₛ Y) => (λₛ (x: X), p • f(-p • x)).
Instance: Support (X →ₛ Y) := fun fs => f_supp fs.
Instance: PermT (X →ₛ Y). Proof. (* ... *) Qed.
Instance: Nominal (X →ₛ Y). Proof. (* ... *) Qed.
\end{coqcode}
\caption{Finitely Supported Functions}
\label{code:supp-func}
\end{code}

The definition of supported functions highlights the necessity of working with setoids in an axiom-free constructive formalization. Defining the sets of supported functions with syntactic equality as the equivalence relation leads the proof of \coqinline{gact_id} from \coqinline{PermT} to a proof of function extensionality, which cannot be proven in Rocq since it is not constructive. Even thought consistent with the Rocq type system, setoids are used instead, to preserve the constructive character of the formalization.

\paragraph{Freshness Theorem} While the \textbf{choose-a-fresh-name principle} states that it is always possible to choose a fresh name (see Section~\ref{sec:constructive-nominal}), the Freshness Theorem states the conditions in which the choice of the fresh variable is irrelevant. This is a new (constructive) formalization, not present in \cite{Choudhury2015}.

\begin{code}
\begin{coqcode}
Context `{Nominal X} (h: Name →ₛ X) (Hp: ∃ a, a # h ∧ a # (h a)).
  
Definition freshF: X := h (fresh (support h)).

Theorem freshness_theorem: ∀ a, a # h → (h a) ≡ freshF.

Corollary freshness_theorem_inj: ∀ a b, a # h → b # h → h a ≡ h b.
\end{coqcode}
\caption{Freshness Theorem}
\label{code:freshness-thm}
\end{code}

Applicability of \coqinline{freshness_theorem} in the formalization, as stated (see Code~\ref{code:freshness-thm}), proved cumbersome to be used and automated, since the theorem is mostly used in proofs to rewrite an argument of the function \coqinline{h} for a new fresh variable. The corollary \coqinline{freshness_theorem_inj}, which is, in fact, a proof of uniqueness for \coqinline{freshF} corresponding to (\ref{def:freshness-handy}) in Section~\ref{sec:constructive-nominal}, showed preferable in the development, since it allowed easier rewrites by explicitly stating the names involved.

\paragraph{Freshness Condition for Binders} The FCB property plays an important role in the derivation of the $\alpha$-structural combinators, but here we want to highlight some aspects of the FCB formalization, in particular when dealing with supported functions. The notation \coqinline{λₛ⟦S⟧ x, t} introduces a new supported function with support \coqinline{S}, \coqinline{x} a set of arguments and \coqinline{t} the body of the function. For the properties \coqinline{f_proper} and \coqinline{f_supp_spec}, the notation introduces holes, \ie{} proof obligations, which can be resolved with the help of the \coqinline{refine} tactic.

\begin{code}
\begin{coqcode}
Context `{Nominal X} `{Nominal Y}.
Context (f: (Name * X) →ₛ Y) (fcb: (∃ a, a # f ∧ ∀ x, a # f (a,x))).

Definition _f: [Α]X →ₛ Y.
refine (
   λₛ⟦support f⟧ (ax: [Α]X), (
      let h: Name →ₛ X := 
        λₛ⟦support (ax.(name)) ∪ support (ax.(term)) ∪ support f⟧ c, 
         (f (c, ⟨ax.(name),c⟩ • ax.(term)))
      in freshF h
   )
).
Proof. (* ... *) Defined.

Lemma fcb_support: support f = support _f.
Lemma fcb_prop (a: Name): a ∉ support f → ∀ x: X, f (a,x) ≡ _f ⟦a⟧x.
\end{coqcode}
\caption{Freshness Condition for Binders}
\label{code:fcb-thm}
\end{code}

%% file: discussion-related.tex
\section{Related and Future Work}\label{sec:discussion-related}

\paragraph{Related Works} 

Presenting the support of some function is not trivial and even in \cite{Urban2008} least support is replaced by some support in some cases. Support cannot be defined from domain/codomain only, \ie\ from the type, of some functions, \eg\ choice functions. A heuristic is used in such cases \cite{Urban2008}, where the function support is considered to be the union of the support of all free symbols in the function definition. For instance, if $f(x) = g(c,x)$ then $supp(f) \subseteq S(g) \cup S(c)$ where $supp$ is the least support and $S$ is some support, \ie\ the heuristic is based on ``some support as an upper bound'' as characterised by (\ref{def:swap-supp}).

Choudhury follows the categorical presentation of nominal sets \cite{Pitts2009} hence their categorical properties were also proved in \cite{Choudhury2015}. We follow an algebraic approach to nominal sets but, apart from that, there is no conceptual difference between our work and his. In fact, several solutions adopted in the former, such as the use of setoids to keep constructiveness, were adopted in the present work. There are also some differences due to the ``ergonomy'' of each proof assistant. For instance, freshness  is a computing artefact in \cite{Choudhury2015} while ours is ``logical''. However, a dependent register with a witness in \coqinline{Set} and an existential formula in \coqinline{Prop} are, in some sense, the same. While the former is used in Agda, where there is no such a distinction as \coqinline{Set} and \coqinline{Prop} in Rocq, the latter is more suitable for the formalization in Rocq. The main advantage in using the Rocq Proof Assistant with a constructive approach is the use of typeclasses combined with the generalized rewriting mechanism, bearing the burden of working with setoids.

More recently, another formalization of nominal sets in Agda was proposed in \cite{Pagano2023}, exploring different representations of finite permutations, instead of restricting it to the usual composition of transpositions, proving their equivalence and defining normalization of composition of transpositions.

\paragraph{Towards $\alpha$-structural principles}

So far, we have developed all basic notions from the nominal theory: permutation types, which provide a uniform interface for handling variables and binders via name permutation; support, which characterizes the set of names that a certain term depends on; freshness, the complementary notion of support, gives us means to decide whether a term depends upon a name; name abstraction, an abstract notion of $\alpha$-equivalence for nominal terms; and supported functions. We have all the tools to tackle syntax with binding variables inside the proof assistant. For instance, we can define the substitution of terms for the $\lambda$-calculus and prove properties such as the composition of substitutions. But still, the process demands extensive boilerplate encodings, not much different than working with de Bruijn indices or with locally nameless terms\footnote{For a comparison among the three approaches in the formalization of the higher-order $\pi$-calculus see \cite{Ambal2020}.}. Ideally, we want to identify terms up to $\alpha$-equivalence, as in a traditional pen-and-paper proof, in which we implicitly choose a term from the $\alpha$-equivalence class satisfying the Barendregt Variable Condition (BVC): all bound names are pairwise distinct and their set is disjoint from the set of free names \cite{Barendregt85}. The $\alpha$-structural principles enable this implicit reasoning \cite{Pitts2006}. Instead of defining a function or proving properties over the recursive/inductive structure of terms, we define/prove it over the terms modulo $\alpha$-equivalence. There is only one full $\alpha$-principle specification presented in \cite{Urban2008}, in the classical setting. Previous attempts in the constructive setting include an axiomatic approach \cite{Aydemir2007} and a specification introduced in \cite{Copello2016}. The former represents a proof-of-concept, where the axioms including the $\alpha$-structural principles would be generated by an external tool from a syntactic and semantic description of the language with binders, while the latter have the principles relying on an extra assumption about the properties of $\alpha$-equivalence class of terms, called $\alpha$-compatibility (see also \cite{Copello2018,CopelloST21}).

Currently, we only have a derivation of an $\alpha$-recursion principle for the $\lambda$-calculus, since its codomain can be any nominal set, while an $\alpha$-induction principle would have \coqinline{Prop} as its codomain.
For any inductively defined type $X$ in the Rocq Calculus of (Co)Inductive Constructions, an associated ``structural principle'' operating over predicates \coqinline{P: X → Type} is generated. Both recursion and induction principles are applications of this general structural principle to predicates with different codomains: \coqinline{Set} for recursions, and \coqinline{Prop} for inductions. And so, both principles consider the same term structure. The same reasoning/term structure could then be used to derive an $\alpha$-structural induction principle, but first we would have to show that \coqinline{Prop} is a nominal set, implicating that the set of the Rocq Core language terms would be a nominal set. 

Instead, we are only developing a nominal theory for any inductively defined \coqinline{X: Set}. It is worth noticing that when a user shows a new inductive type \coqinline{X: Set} to be nominal, we have a "confined" signature to work over it. All constructors can be seen as uninterpreted functions over \coqinline{X}, and can be proved nominal with an appropriate definition of support. The semantics assigned to its constructors will come from all user-defined functions over \coqinline{X}, which can be shown to be Nominal if both domain and codomain are also nominal. As an ongoing work, we are investigating some alternatives for an induction principle, with the induction principle from \cite{Copello2016,CopelloST21} posing the best alternative so far.